# Spatial Effects in Convergence of Portuguese Product

**Vitor João Pereira Domingues Martinho**

Unidade de I&D do Instituto Politécnico de Viseu
Av. Cor. José Maria Vale de Andrade
Campus Politécnico
3504 - 510 Viseu
**(PORTUGAL)**
e-mail: vdmartinho@esav.ipv.pt

**ABSTRACT**

This study analyses, through cross-section estimation methods, the influence of spatial effects in the conditional product convergence in the parishes' economies of mainland Portugal between 1991 and 2001 (the last year with data available for this spatial disaggregation level). To analyse the data, Moran's I statistics is considered, and it is stated that product is subject to positive spatial autocorrelation (product develops in a similar manner to product in neighbouring regions). Taking into account the estimation results, it is stated that there are not indications of convergence (the population is in the littoral of Portugal) and it can be seen that spatial spillover effects, spatial lag (capturing spatial autocorrelation through a spatially lagged dependent variable) and spatial error (capturing spatial autocorrelation through a spatially lagged error term) condition the convergence of product of Portuguese parishes in the period under consideration (1)(Martinho, 2011).

**Keywords:** Spatial Econometrics; Product Convergence; Portuguese Context.

**1. INTRODUCTION**

There are some known studies concerning conditional convergence with spatial effects. (2)Fingleton (2001), for example has found spatial correlation at the level of productivity when, using data from 178 regions of the European Union, he introduced spillover effects in a model of endogenous growth. (3)Abreu et al. (2004) have investigated the spatial distribution of the rates of total productivity growth of factors using exploratory analyses of spatial data and other techniques of spatial econometrics. The sample consists of 73 countries and covers the period from 1960 to 2000. They have found significant spatial correlation in the rates of total factor productivity growth, indicating that high and low values tend to concentrate in space, forming the so-called "clusters". They have also found high indications of positive spatial autocorrelation at the level of the total factor productivity, which has increased throughout the period of 1960 to 2000. This result could indicate a tendency to clustering with time. (4)NijKamp (2007) analyses spatial disparities, differences in regional growth and productivity and discuss their driving forces. In this paper starting from traditional regional growth theory, it introduces next findings from location and agglomeration theory, including infrastructure and network modelling, with a particular emphasis on spatial accessibility. Innovation, entrepreneurship and knowledge are addressed, too, and interpreted as critical success conditions for modern regional development. Next, elements from endogenous growth theory and the new economic geography are introduced.

There is, on the other hand, a variety of studies analysing conditional product convergence with spatial effects. Armstrong has defended that the evidence of convergence across European countries as mentioned by Barro and Sala-i-Martin is due to the omission of spatial autocorrelation in their analysis and bias resulting from the selection of European regions. Following on, (5)Sandberg (2004), for example, has examined the hypothesis of absolute and conditional convergence across Chinese provinces in the period from 1985 to 2000 and found indications that there had been absolute convergence during the periods of 1985 to 2000 and 1985 to 1990. He has also found evidence that conditional convergence had been seen in the sub-period of 1990 to 1995, with signs of spatial dependency across adjacent provinces. (6)Arbia et al. (2004) have studied the convergence of gross domestic product per capita among 125 regions of 10 European countries from 1985 to 1995, considering the influence of spatial effects. They concluded that the consideration of spatial dependency considerably improved the rates of convergence. (7)Lundberg (2004) has tested the hypothesis of conditional convergence with spatial effects between 1981 and 1990 and, in contrast to previous results, found no clear evidence favouring the hypothesis of conditional convergence. On the contrary, the results foresaw conditional divergence across municipalities located in the region of Stockholm throughout the period and for municipalities outside of the Stockholm region during the 1990s.

This study seeks to test conditional product convergence (using as a proxy the population, following the new economic geography theory, because the persons are potential workers and consumers, so where we have more persons we have more product) for the parishes economies of mainland Portugal from 1991 to 2001, through techniques of cross-section spatial econometrics. To do so, this study is structured in five parts: after this introduction, there follows the second part where the models considered are explained; in the third part the data based on techniques of spatial econometrics developed to explore spatial data are analysed; in the fourth part the estimates drawn up are presented and in the fifth the main conclusions obtained after this research are highlighted.

## 2. MODEL OF CONDITIONAL CONVERGENCE WITH SPATIAL EFFECTS

Bearing in mind the theoretical considerations about conditional convergence, what is presented next is the model used to analyse conditional product convergence with spatial effects, at parishes' level in mainland Portugal:

$$(1/T)\log(P_{it}/P_{i0}) = \alpha + \rho W_{ij} p_{it} + b \log P_{i0} + \varepsilon_{it}, \text{ with } \alpha > 0 \text{ e } \beta < 0 \qquad (1)$$

In this equation (1) P is the product, p is the rate of growth of product in various regions, W is the matrix of distances, b is the convergence coefficient, $\rho$ is the autoregressive spatial coefficient (of the spatial lag component) and $\varepsilon$ is the error term (of the spatial error component, with, $\varepsilon = \lambda W \varepsilon + \xi$). The indices i, j and t, represent the regions under study, the neighbouring regions and the period of time respectively.

## 3. DATA ANALYSIS

The data referring to population were obtained in the censos (population data) of the National Statistics Institute. To carry out the cross-section estimations, the GeoDa[1] software was used (8)(Anselin 2010).

What follows is an analysis to identify the relationship between the dependent and independent variable and the existence of spatial autocorrelation by using Moran Scatterplots for over all spatial autocorrelation and LISA Maps for local spatial autocorrelation.

The following Scatterplot, (showing the relation between the growth of product and initial product) presented below, allow for an analysis of product convergence for Portuguese parishes, since 1991 to 2001.

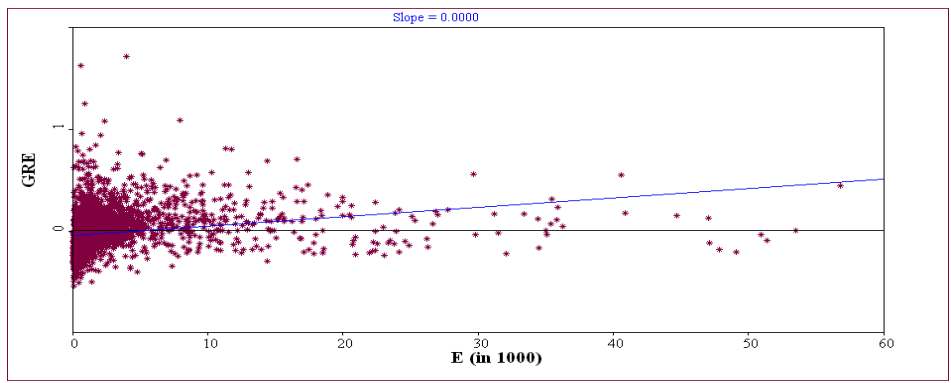

**Figure 1:** Scatterplot of absolute convergence of product

Note: GRE = Growth rate product;
E = Initial product.

Analysing the figure above we see now absolute convergence for the product of the Portuguese parishes.

The Moran Scatterplot (showing the relationship between the dependent variable and the spatially lagged dependent variable) which are presented below, show Moran's I statistical values for the 4052 parishes of the mainland Portugal from 1991 to 2001. The matrix $W_{ij}$ used is the matrix of the distances between the regions up to a maximum limit of 97 Km. This distance appeared to be the most appropriate to the reality of Portuguese context, given the signs of spatial autocorrelation encountered, (with an analysis of the data, bearing in mind namely Moran's I statistics, and with the estimation results carried out) in the analysis of robustness and behaviour of the various matrices of distance when considering alternative possibilities of maximum distances. Whatever the case, the choice of the best limiting distance to construct these matrices is always complex.

---
[1] Available at http://geodacenter.asu.edu/

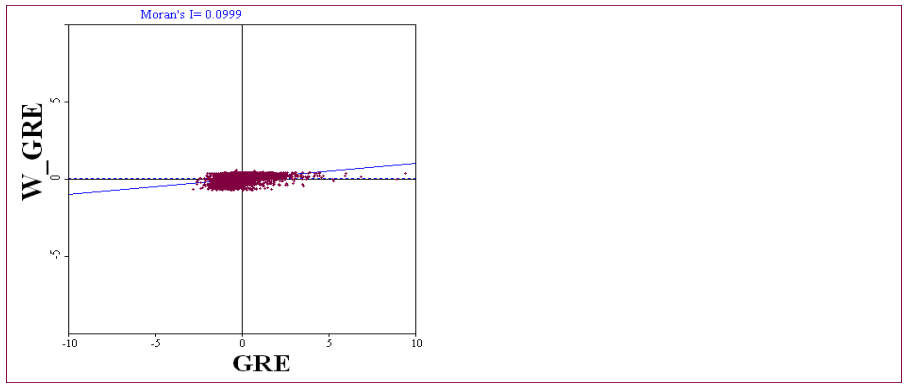

**Figure2:** Moran Scatterplot of product

Note: W-GRE = Spatially lagged growth rate product;
GRE = Growth rate product.

An analysis of the Moran Scatterplot shows that there is some, but weak, spatial autocorrelation for the dependent variable.

Figure 3 analyses the existence of local spatial autocorrelation with LISA Map, investigated under spatial autocorrelation and its significance locally. The parishes with high-high and low-low values, correspond to the regions with positive spatial autocorrelation and with statistical significance. The regions with high-low and low-high values are outliers with negative spatial autocorrelation.

Analysing the LISA Cluster Map above we see spatial autocorrelation high-high for the littoral and spatial autocorrelation low-low for the interior of Portugal.

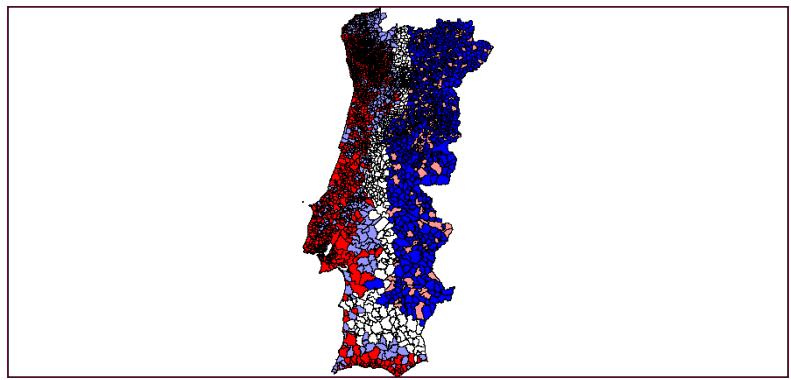

**Figure 3**: LISA Cluster Map of product

Note:

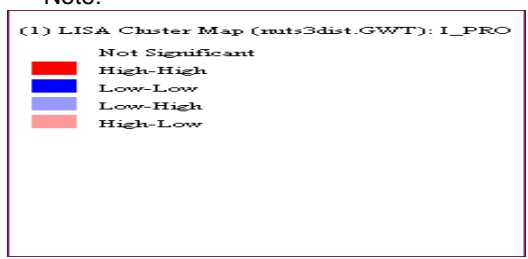

## 4. EMPIRICAL EVIDENCE FOR CONDITIONAL CONVERGENCE OF PRODUCT, CONSIDERING THE POSSIBILITY THAT THERE ARE SPATIAL EFFECTS

What follows is the presentation of empirical evidence of the existence of conditional product convergence based on cross-section estimates. The cross-section estimates were carried out using the Least Square (OLS) method and the Maximum Likelihood (ML) method. The use of the Maximum Likelihood method is based on the fact that the associated authors advise using it when there is the presence of spatial spillover effects.

This part of the study will examine the procedures of specification by (9)Florax et al. (2003) and will firstly examine through OLS estimates, the relevance of proceeding with estimate models with spatial lag and spatial error components with recourse to LM specification tests.

The results concerning OLS estimates of conditional convergence with tests of spatial specification are present in Table 1, which follows.

**Table 1:** OLS estimation results for the equation of absolute convergence with spatial specification tests

$$(1/T)\log(P_{it}/P_{i0}) = \alpha + b\log P_{i0} + \varepsilon_{it}$$

| Constant | Coef. b | JB | BP | KB | M'I | LM$_l$ | LMR$_l$ | LM$_e$ | LMR$_e$ | $\bar{R}^2$ | N.O. |
|---|---|---|---|---|---|---|---|---|---|---|---|
| -0.052* (-16.333) | 0,001* (15.639) | 1.819 | 1.012 | 1.052 | 3.074* | 92.164* | 3.764* | 187.805* | 98.405* | 0.567 | 4050 |

**Note: JB, Jarque-Bera test; BP, Breusch-Pagan test; KB, Koenker-Bassett test: M'I, Moran's I; LM$_l$, LM test for spatial lag component; LMR$_l$, robust LM test for spatial lag component; LM$_e$, LM test for spatial error component; LMR$_e$, robust LM test for spatial error component;R$^2$, coefficient of adjusted determination; N.O., number of observations; *, statistically significant to 5%; **, statistically significant to 10%.**

This confirms to what has been previously seen in the data analysis, or, in other words, product diverged in Portugal between 1991 and 2001. There are not indications of heteroskedasticity, according to the BP and KB tests. Convergence/divergence in the product will be conditioned by spillover and spatial error effects according to the LM tests.

Table 2 presents the results of the estimates of spillover and spatial error effects.

**Table 2:** ML estimation results for the equation of conditional convergence with spatial effects

$$(1/T)\log(P_{it}/P_{i0}) = \alpha + \rho W_{ij} p_{it} + b\log P_{i0} + \varepsilon_{it}$$

| Constant | Coefficient | Spatial coefficient | Breusch-Pagan | $\bar{R}^2$ | N.Observations |
|---|---|---|---|---|---|
| -0.267 (-0.978) | 0.001* (10.227) | 0.990* (150.612) | 1.958 | 0.653 | 4050 |

**Note: *, statistically significant to 5%; **, statistically significant to 10%; ***, spatial coefficient of the spatial error model.**

The convergence coefficient are more or less the same, but the spatial coefficient confirm the existence of spatial autocorrelation.

### 5. CONCLUSIONS

This study has sought to test the convergence of product across the 4052 parishes of mainland Portugal in the period of 1991 to 2001, with spillover, spatial lag and spatial error effects. To do so, data analysis and cross-section estimates have been carried out with the OLS and ML estimation methods, following the specification procedures indicated by Florax et al. (2003) who suggest that models are first tested with the OLS method, to test which is the better specification (spatial lag or spatial error) and then the spatial lag or spatial error is estimated with the ML method.

Considering the analysis of the data and the estimation results previously carried out, it can be seen that product is subject to positive spatial autocorrelation and that the product diverged between 1991 and 2001 in Portugal. This is a preoccupant situation, because we are the population all in the littoral and no people in the interior.